\documentclass[seceq]{ptptex}

\usepackage{graphicx}

\newcommand{\feyn}[1]{
  \setbox0=\hbox{\ensuremath{#1}}
  \hbox to\wd0{\hbox to0pt{\hbox to\wd0{\hss/\hss}\hss}\box0}}

\title{Bose-Einstein condensation of diquark molecules in
three-flavor quark matter}

\author{
Masakiyo \textsc{Kitazawa}$^{1,}$\footnote{ e-mail address:
masky@yukawa.kyoto-u.ac.jp},  
Dirk H.\ \textsc{Rischke}$^{2,}$\footnote{ e-mail address:
drischke@th.physik.uni-frankfurt.de}
and
Igor A.\ \textsc{Shovkovy}$^{3,}$\footnote{ e-mail address:
I-Shovkovy@wiu.edu} 
}

\inst{
$^1$ RIKEN-BNL Research Center, Brookhaven National Laboratory,
Upton, NY 11973, USA\\
$^{2}$ Institut f\"ur Theoretische Physik
and Frankfurt Institute for Advanced Studies,
J.W.\ Goethe-Universit\"at,
D-60438 Frankfurt am Main, Germany\\
$^{3}$ Department of Physics, Western Illinois University,
Macomb, IL 61455, U.S.A.
}

\abst{
We study the phase diagram of strongly interacting 
matter with three quark flavors at low and intermediate
densities and non-zero temperatures in the framework of
an NJL-type model with four-point interactions. 
At large densities, when the interactions are weak due to
asymptotic freedom, quarks form loosely bound Cooper pairs.
However, when the density decreases, interactions 
become stronger and quark Cooper pairs transform smoothly
into tightly bound diquark molecules. We
find that such molecules are stable at low density and
temperature and that they dissociate above a temperature
$T_{\rm diss}$ of the order of the chiral phase transition
temperature $T_c \sim 170$ MeV.
We also explore the conditions under which these
molecules undergo Bose-Einstein condensation (BEC).
We find that BEC is only possible if we
increase the attractive interaction in the diquark channel
to (probably unrealistically) large values.
}

\begin{document}

\maketitle

\section{Introduction}
\label{rischke_intro}

At asymptotically large densities, due to 
asymptotic freedom of QCD \cite{Gross:1973id}, 
the interaction between quarks becomes arbitrarily weak.
Then, single-gluon exchange constitutes the dominant contribution
to the quark-quark interaction.
Single-gluon exchange is attractive in the 
color-antitriplet channel. Thus, we expect the formation 
of quark Cooper pairs which destabilize the Fermi 
surface \cite{Bardeen:1957mv} and lead to color 
superconductivity \cite{Bailin:1983bm}, 
see \cite{reviews} for reviews.

At sufficiently large densities, the coupling constant
is sufficiently small so that 
color superconductivity can be analyzed
rigorously in the framework of QCD, using resummation
techniques based on perturbative methods and power counting
\cite{Son:1998uk,Schafer:1999jg,Pisarski:1999tv,Hong:1999fh}.
This treatment is the analogue of weak-coupling
BCS theory in condensed matter physics.
The zero-temperature gap parameter turns out to be 
parametrically small in the coupling constant, 
$\phi_0 \sim \mu \, \exp(-1/g)$, where $\mu$ is the quark
chemical potential. This, in turn, leads to
``large'' Cooper pairs, i.e., the quark correlation length 
$\xi \sim \phi_0^{-1} \sim \mu^{-1} \exp(1/g)$ 
is parametrically larger than the interparticle distance 
$\sim \mu^{-1}$ \cite{AHI02}.

When the density decreases, however, the strength of the
quark-quark interaction and, thus, the gap parameter, increase,
and the correlation length decreases. Let us note that
a signature of strong correlations in
the normal phase is the appearance of a so-called pseudogap
in the vicinity of the transition to the superconducting phase
\cite{KKKN02,KKKN04}. When the correlation length
$\xi \lesssim \mu^{-1}$, quark Cooper pairs should be 
regarded as tightly bound diquark molecules.
These molecules are stable at all temperatures below
their so-called dissociation temperature 
$T_{\rm diss}$ \cite{NSR85}. Above $T_{\rm diss}$, they
decay into the two quarks constituting the molecule.
At small temperatures, 
diquark molecules may undergo Bose-Einstein condensation (BEC)
\cite{Legg80,NSR85,AHI02,Nishida:2005ds,Nawa:2005sb,Abuki:2006dv,Rezaeian:2006yj,Deng:2006ed,Ebert:2006tc,Sun:2007fc}.

In these proceedings, we investigate bound 
diquark states and the possibility that they undergo
BEC in the phase diagram of the quark matter. We use
an NJL-type model with four-quark interactions.
The strength of the attractive diquark interaction 
in the color-antitriplet
channel is regarded as a free parameter.
We show that bound diquark molecules appear at low densities and
temperatures for all values of the diquark coupling strength
studied here. We also find that BEC of diquarks occurs for
(probably unrealistically) large values of the diquark coupling.

The remainder of this work is organized as follows.
In Sec.\ \ref{rischke_form} we introduce the model and 
the formalism in order to study diquark correlations.
In Sec.\ \ref{rischke_num} we present our numerical results.
We conclude our work in Sec.\ \ref{rischke_conc}.
Our units are $\hbar = k_B = c = 1$, the metric tensor
is $g^{\mu \nu} = {\rm diag}(+,-,-,-)$.

\section{Model and Formalism}
\label{rischke_form}

In this work, we employ an NJL-type model Lagrangian for
three quark flavors,
\begin{eqnarray}
\mathcal{L}& =& \bar \psi \, ( i \partial \hspace{-0.5em} / 
- \hat{m} \, ) \psi 
+G_S \sum_{a=0}^8 \left[ \left( \bar \psi \lambda_a \psi 
\right)^2 
+ \left( \bar \psi i \gamma_5 \lambda_a \psi \right)^2 \right] 
\nonumber \\
& +& G_D \sum_{\gamma,c} \left[\bar{\psi}_{\alpha}^{a} \, 
i \gamma_5 \epsilon^{\alpha \beta \gamma}
\epsilon_{abc} \, (\psi_C)_{\beta}^{b} \right] \left[ 
(\bar{\psi}_C)_{\rho}^{r} \, i \gamma_5
\epsilon^{\rho \sigma \gamma} \epsilon_{rsc} \, 
\psi_{\sigma}^{s} \right] ,
\label{rischke_eq:Lagrangian}
\end{eqnarray}
where the quark field $\psi_{\alpha}^{a}$ has color,
$a=r,g,b$, and flavor, $\alpha=u,d,s$, indices. 
The current quark mass matrix is given by 
$\hat{m} = {\rm diag}_{f}(m_u, m_d, m_s)$, and
$\lambda_a$ are (twice) the generators of $U(3)$.
The charge-conjugate spinors are 
$\psi_C = C \bar \psi^T$ and $\bar \psi_C = \psi^T C$, 
where $C=i\gamma^2 \gamma^0$ 
is the charge conjugation matrix. 
A sum over doubly appearing upper and lower
indices in color and flavor space is implied (but not
if both are upper or lower indices).

In the Lagrangian (\ref{rischke_eq:Lagrangian}), the
terms proportional to $G_S$ are
quark-antiquark four-point interactions in the scalar
and pseudoscalar channel, respectively. The
terms proportional to $G_D$ parametrize the diquark four-point
interaction in the color-antitriplet, flavor-antitriplet
channel. For one-gluon exchange in QCD, 
this channel is attractive and thus leads to 
color superconductivity. Note that the diquark term
can also be obtained from the quark-antiquark term
by a Fierz transformation; in this case, the diquark
coupling strength is fixed, $G_D = 0.75\, G_S$
\cite{Vanderheyden:1999xp}.
For the sake of simplicity, we neglect the effect of the
$U(1)_A$ anomaly, so there is no 
t'Hooft-type six-point interaction term 
in Eq.~(\ref{rischke_eq:Lagrangian}).

In mean-field approximation, the thermodynamic potential is
\begin{equation}
  \Omega = \sum_{c=1}^{3} \frac{|\Delta_c|^2}{4G_D}+
   \sum_{\alpha=1}^{3}\frac{(M_\alpha-m_\alpha)^2 }{8G_S}
-\frac{T}2 \sum_n \int\frac{d^3{\bf p}}{(2\pi)^3}\,
  {\rm Tr}_{D,f,c}
  \ln \left[S^{-1}(i\omega_n,{\bf p})\right], 
\label{rischke_eq:Omega}
\end{equation}
where 
\begin{eqnarray}
  M_\alpha &=& m_\alpha - 4G_S 
            \langle \bar\psi_\alpha \psi_\alpha \rangle , \\
  \Delta_c&=& 2G_D \langle \bar{\psi}_\alpha^a \, 
           (P_c)_{ab}^{\alpha\beta}\, \psi_\beta^b \rangle,
\end{eqnarray}
are the constituent quark masses and the gap parameters for 
color superconductivity, respectively, 
with $(P_c)_{ab}^{\alpha\beta}
 = i\gamma_5 {\epsilon^{\alpha\beta}}_c \epsilon_{ abc }$.
The $72\times 72$ Nambu-Gor'kov propagator is defined by
\begin{equation}
   S^{-1}(i\omega_n,{\bf p})= \left( \begin{array}{cc}
  \feyn{p}+\mu\gamma_0-{\bf M}& \sum_c
   P_c\Delta_c  \\
  \sum_c \bar{P}_c\Delta_c & 
  {^t\feyn{p}-\mu}\gamma_0+{\bf M}
  \end{array} \right),
\end{equation}
with $\feyn{p}=i\omega_n\gamma_0-{\bf p}\cdot{\bf \gamma}$.
Here, $\omega_n=(2n+1) \pi T$, $n = 0, \pm 1, \pm 2, \ldots$, 
are the fermionic Matsubara frequencies.
In this first exploratory study of diquark molecules and
BEC, we assume the quark chemical potential $\mu$ to have 
a common value for all flavors. In application to compact
stellar objects, this changes due to the conditions of
overall electric (and color) neutrality, as well as
$\beta$ equilibrium. In application to heavy-ion collisions,
this is also different, as (for isospin-symmetric nuclei)
$\mu_u = \mu_d = \mu$, while $\mu_s = 0$.

The physical values of the variational 
parameters $\Delta_c$ and $M_\alpha$
satisfy the gap equations
\begin{equation} 
\frac{ \partial\Omega }{ \partial \Delta_c } = 0 \quad 
{\rm and} \quad
\frac{ \partial\Omega }{ \partial M_\alpha } = 0.
\end{equation}
In mean-field approximation, the phase transition from
the color-superconducting to the normal phase is
of second order. This means that the order parameter
for condensation, $\Delta_c$, goes to zero smoothly as
a function of temperature and
the critical temperature for the transition can be
determined from
\begin{equation}
\left.
\frac {1}{\Delta_c} \frac{ \partial\Omega }{ \partial\Delta_c }
\right|_{\Delta_c=0} = 0.
\label{rischke_eq:crit}
\end{equation}

Since up and down flavors are treated symmetrically
in our model, we assume 
$M_u=M_d$ and $\Delta_1=\Delta_2$.
Because of explicit chiral symmetry breaking by
a nonzero current quark mass,
$\langle \bar\psi_\alpha \psi_\alpha \rangle$ is nonzero 
for all $T$ and $\mu$.
In the following, we refer to the phase with 
$\Delta_3 \ne 0$ and $\Delta_{1,2}=0$ as the 2SC phase,
and $\Delta_3 \ne 0$ and $\Delta_{1,2} \ne 0$ as the CFL phase
\cite{reviews}.
The phase with $\Delta_1=\Delta_2=\Delta_3 = 0$ corresponds
to unpaired quark matter.

At nonzero temperature, the order parameters
$\Delta_c$ and $M_\alpha$ fluctuate around 
their mean-field values.
The propagation of these fluctuations in unpaired quark matter
is characterized by the retarded propagator
\begin{eqnarray}
D_c^R ( {\bf x},t ; {\bf x}',t' ) 
& = & -i \theta(t-t')\langle 
[\bar{\psi}( {\bf x},t ) P_c \psi_C( {\bf x},t ), 
\bar{\psi}_C ( {\bf x}',t' ) P_c \psi({\bf x}',t')]
\rangle \nonumber \\
& = & \int \frac{d\omega d^3 {\bf k}}{(2\pi)^4} \,
D_c^R ( \omega ,{\bf k})\,
{\rm e}^{ -i\omega (t-t') + i{\bf k} \cdot ({\bf x}-{\bf x}')},
\label{rischke_eq:D^R}
\end{eqnarray}
where $c=1,2,$ or $3$ denotes the down-strange, up-strange,
or up-down diquark field, respectively.
In the random phase approximation,
the diquark propagators are given by
\begin{equation}
  D_c^R(\omega,{\bf p})= \frac{1}{2}\,
  \frac{Q_c^R(\omega,{\bf p})}{1+G_D Q_c^R(\omega,{\bf p})},
\end{equation}
where 
$Q_c^R(\omega,{\bf p})$ is the one-loop quark-quark 
polarization function. For imaginary energies $\omega = i \nu_n$,
where $\nu_n=2n\pi T$, $n=0, \pm 1, \pm 2, \ldots$, are the
bosonic Matsubara frequencies, it is given by
\begin{equation}
  \mathcal{Q}_c(i\nu_n,{\bf p})
  = 2T\sum_m \int \frac{d^3 {\bf q}}{(2\pi)^3}
  |{\epsilon_c}^{\beta\gamma}|
    {\rm Tr}_{D,c} [ \mathcal{G}_\beta(i\omega_m,{\bf q})
  \mathcal{G}_\gamma(i\nu_n+i\omega_m,{\bf p}+{\bf q}) ],
\label{rischke_eq:Q}
\end{equation}
with the trace taken over Dirac and color indices.
Here, ${\cal G}_\alpha (i\omega_n,{\bf p}) =
[ ( i\omega_n+\mu )\gamma_0 -{\bf p}\cdot{\bf \gamma} 
  - M_\alpha ]^{-1}$
are the Matsubara Green's functions for quarks of flavor $\alpha$.
Substituting these Green's function into Eq.~(\ref{rischke_eq:Q})
and taking the analytic continuation 
$Q_c^R(\omega ,{\bf p})={\cal Q}_c(i\nu_n,{\bf p})
|_{i\nu_n\to \omega+i\eta}$,
we obtain
\begin{eqnarray}
Q^R_c(\omega,{\bf p})
&=&
-2 \sum_{\beta,\gamma} |\epsilon_{c\beta\gamma}|
\int \frac{ d^3{\bf q} }{ (2\pi)^3 } \sum_{s,t=\pm}
st \, \frac{ ( sE_\beta + tE_\gamma )^2 - |{\bf p}|^2 
- \delta M_c^2 }{ E_\beta E_\gamma }\nonumber \\
&  & \hspace*{4.2cm}
\times\; \frac{ f( tE_\gamma-\mu ) - f( -sE_\beta+\mu ) }
{ \omega + 2\mu - sE_\beta - tE_\gamma + i\eta }\,,
\end{eqnarray}
where $E_\beta = \sqrt{ |{\bf q}-{\bf p}|^2 + M_\beta^2 }$, 
$E_\gamma = \sqrt{ |{\bf q}|^2 + M_\gamma^2 }$,
$\delta M_c = | M_\beta - M_\gamma |$ and 
$ f(E) = [ \exp(E/T) + 1 ]^{-1}$ is the Fermi-Dirac 
distribution function.
The imaginary part of $Q^R_c(\omega,{\bf p})$ denotes
the difference of decay and production rates of the 
diquark field.
At ${\bf p}=0$, it is given by
\begin{eqnarray}
{\rm Im} Q_c^R (\omega, 0 )
& = & 2\pi \sum_{\beta, \gamma}|\epsilon_{c\beta\gamma}| 
\int \frac{ d^3{\bf q} }{ (2\pi)^3 }
\frac{ ( \omega+2\mu )^2 - \delta M_c^2 }{ E_\beta E_\gamma } 
\nonumber \\
&   \times& \left\{
-\left[ ( 1-f_\beta^+ )( 1-f_\gamma^+ ) - f_\beta^+ 
f_\gamma^+ \right]
\delta( \omega + 2\mu - E_\beta - E_\gamma ) \right.
\nonumber \\
& & \hspace*{0.25cm} +\left[ ( 1-f_\beta^- )( 1-f_\gamma^- ) - 
f_\beta^- f_\gamma^- \right]
\delta( \omega + 2\mu + E_\beta + E_\gamma )
\nonumber \\ 
& & \hspace*{0.25cm} -\left[ f_\beta^- ( 1-f_\gamma^+ ) - 
( 1 - f_\beta^- ) f_\gamma^+ \right]
\delta( \omega + 2\mu + E_\beta - E_\gamma )
\nonumber \\ 
& & \hspace*{0.22cm} \left. -\left[ f_\beta^+ ( 1-f_\gamma^- ) - 
( 1 - f_\beta^+ ) f_\gamma^- \right]
\delta( \omega + 2\mu - E_\beta + E_\gamma )\right\} \,,
\label{rischke_eq:ImQR_k0}
\end{eqnarray}
where $f_\alpha^\pm = \{ \exp[  ( E_\alpha \mp \mu )/T ] 
+ 1 \}^{-1}$.
The first (second) term in curly brackets
corresponds to the decay of a diquark into two quarks 
(anti-quarks) and assumes nonzero values 
for $\omega > 2\bar{M}_c-2\mu$ and $\omega < -2\bar{M}_c-2\mu$,
with $\bar{M}_c = ( M_\beta + M_\gamma ) / 2 $.
The third and fourth terms represent Landau damping of
a diquark.
They are nonzero for $ -\delta M_c -2\mu < \omega < 
\delta M_c-2\mu $.
The numerical results show that the latter processes
do not affect the stability of diquark excitations
in our model, since the pole of the diquark field never appears
around energies where Landau damping occurs.

The poles of the diquark propagator $D_c^R$ are
determined by solving
${D_c^R(\omega,{\bf p})}^{-1}=0$ or, equivalently,
\begin{equation}
1 + G_D Q^R_c(\omega,{\bf p}) =0.
\label{rischke_eq:pole}
\end{equation}
Setting $\omega=|{\bf p}|=0$ in this equation,
one can show that Eq.~(\ref{rischke_eq:pole}) 
is equivalent
to the condition Eq.~(\ref{rischke_eq:crit}).
This means that there exists a pole of the diquark propagator
for zero energy and momentum at the critical temperature
$T_c$ of a second-order phase transition. This fact is
known as the Thouless criterion \cite{Thou}.
Above $T_c$, the pole moves continuously to the fourth quadrant
of the complex $\omega$-plane.
The corresponding mode is called the soft mode.
If $\bar{M}_c<\mu$ at $T=T_c$, $\omega=0$ is 
in the continuum and the soft mode has a decay width 
above $T_c$ \cite{KKKN02}. 
On the other hand, if $\bar{M}_c>\mu$, 
the soft mode does not have
a decay width and the pole stays on the real axis. 
Then, this mode is nothing but a 
stable diquark molecule \cite{NSR85}.
As $T$ increases, the pole will move along the
real axis until, at the dissociation 
temperature $T^c_{\rm diss}$, it eventually arrives at
the threshold of the decay process into two quarks
$\omega_{\rm thr}^c=2(\bar{M}_c-\mu)$ . (Note that
$\bar{M}_c - \mu$ is the energy required to put one additional
quark into the system. If the diquark energy exceeds twice this
value, the diquark will decay.)
The dissociation 
temperature $T^c_{\rm diss}$ is determined by solving
\begin{equation}
1 + G_D Q_c^R ( 2\bar{M}_c-2\mu,0 ) = 0.
\end{equation}
If stable diquarks are formed above $T_c$, 
it is natural to associate the superfluid phase below $T_c$
with BEC of these diquark molecules.
In the following, we regard the region of a
color-superconducting
phase where $\mu < \bar{M}_c$ is satisfied
as a Bose-Einstein condensed phase \cite{NSR85}.
Note, however, that this is just a rough estimate to separate
BEC and BCS regimes; these two limits are connected 
continuously and there is no sharp phase boundary \cite{NSR85}.

\section{Numerical Results}
\label{rischke_num}

\begin{figure}
\begin{center}
\includegraphics[width=.8\textwidth]{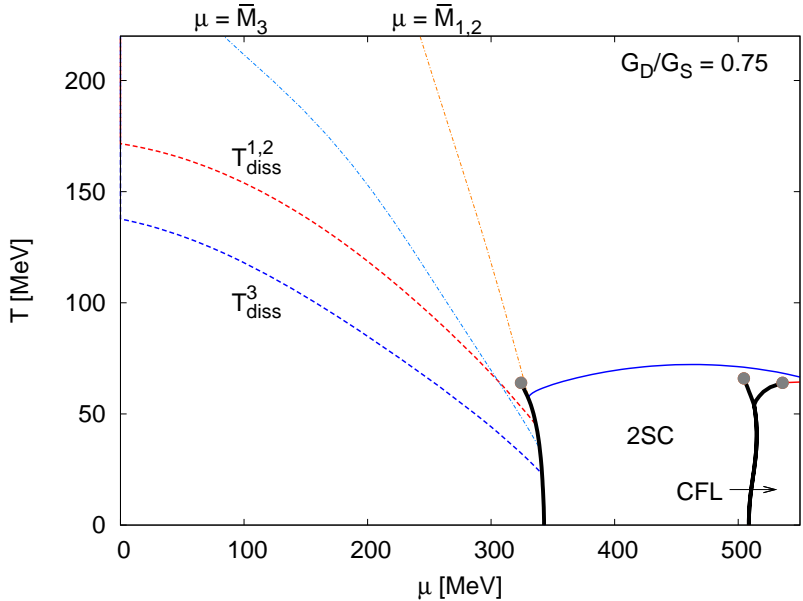} 
\caption{
The phase diagram in the $T$-$\mu$ plane for $G_D/G_S=0.75$.
Bold and thin solid lines represent first- and 
second-order phase transitions, respectively,
while dotted lines denote the dissociation temperature 
of diquark molecules formed of up and down quarks, 
$T^3_{\rm diss}$, and of up and strange as well as down and
strange quarks, $T^{1,2}_{\rm diss}$, respectively.
The condition $\mu=\bar{M}_c$ is denoted by 
dash-dotted lines.
}
\label{rischke_fig:phased1}
\end{center} 
\end{figure}

In this section, we present the phase diagram 
in the $T$-$\mu$ plane
for several values of the diquark coupling constant $G_D$.
In Fig.~\ref{rischke_fig:phased1},
we show the phase diagram for $G_D/G_S = 0.75$,
which is the canonical value arising from
a Fierz transformation of the NJL-type four-point 
interaction \cite{Vanderheyden:1999xp,Buballa:2001gj}.
One observes two color-superconducting phases, 
the 2SC and CFL phases, at high $\mu$ and low $T$.
At $T=0$, these phases are separated by a first-order 
phase transition;
the first-order transition terminates at nonzero temperature.
Also shown in Fig.~\ref{rischke_fig:phased1} are
the dissociation temperatures for stable diquark
molecules; below the curve labelled $T^3_{\rm diss}$, 
bound states
of up and down quarks are stable, and below
$T^{1,2}_{\rm diss}$, bound states of up with strange and
down with strange quarks are stable.

As discussed in the previous section, BEC of diquark molecules
requires that $\mu <\bar{M}_c$ inside a color-superconducting
phase. In Fig.~\ref{rischke_fig:phased1}, 
we show where $\mu = \bar{M}_c$;
the regions to the left of these lines satisfy $\mu<\bar{M}_c$.
One sees that these lines terminate at the first-order 
transition between normal and superconducting phase
and that $\mu < \bar{M}_c$ is not satisfied inside
a color-superconducting phase.
Therefore, there is no BEC for $G_D/G_S = 0.75$.

\begin{figure}
\begin{center}
\includegraphics[width=.8\textwidth]{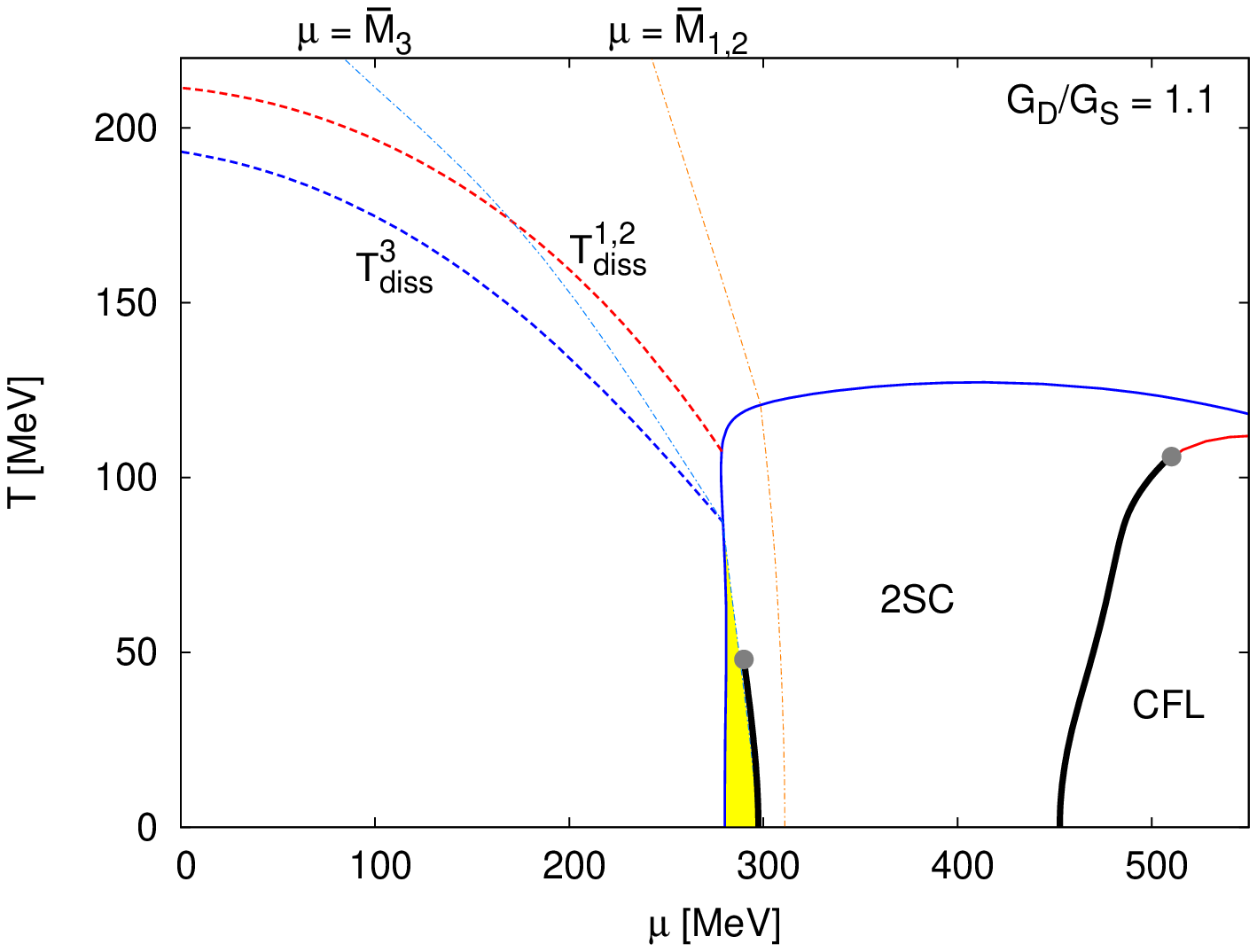} 
\caption{
The phase diagram in the $T$-$\mu$ plane
for $G_D/G_S=1.1$.
}
\label{rischke_fig:phased2}
\end{center} 
\end{figure}

In Fig.~\ref{rischke_fig:phased2}, we show the phase diagram
for $G_D/G_S=1.1$.
We see that the regions of the 2SC and CFL phases 
expand towards lower $\mu$ and higher $T$.
Now there appears BEC inside the 2SC phase,
namely where $\mu <\bar{M}_3$, 
shown by the shaded area in Fig.~\ref{rischke_fig:phased2}.
One also sees that the dissociation temperatures 
$T^c_{\rm diss}$ increase with $G_D$.
The values of $T^c_{\rm diss}$ at $\mu=0$ are 
comparable to or even larger than
the critical temperature of the QCD phase transition
as determined by lattice QCD calculations,
$T_{c, \rm LQCD} \simeq 170$ MeV \cite{Lat_Tc}.
Therefore, bound diquarks can survive even
in the quark-gluon plasma phase if the diquark coupling is 
sufficiently large.

The other interesting feature shown in 
Fig.~\ref{rischke_fig:phased2}
is the fate of the first-order phase transition.
One sees that the corresponding transition line
terminates at smaller temperatures for $G_D/G_S=1.1$ 
than for $G_D/G_S=0.75$. 
This result can be interpreted as a result of the
interplay between chiral symmetry breaking and color
superconductivity.
As far as we have checked, the endpoint of the 
first-order transition smoothly approaches $T=0$
as $G_D$ increases. \footnote{
We do not observe a new endpoint at a lower
temperature, found in Ref.\ \cite{KKNNHTYB}.}

\begin{figure}
\begin{center}
\includegraphics[width=.8\textwidth]{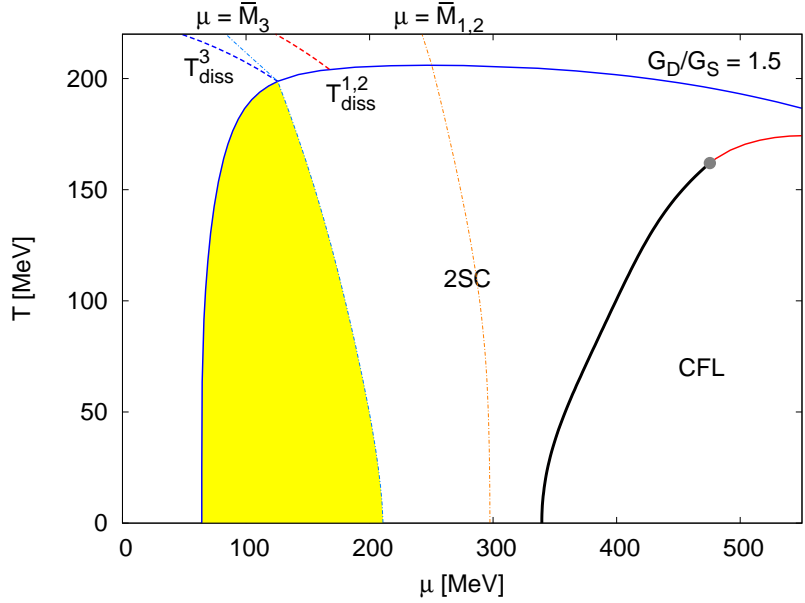} 
\caption{
The phase diagram in the $T$-$\mu$ plane
for $G_D/G_S=1.5$.
}
\label{rischke_fig:phased3}
\end{center} 
\end{figure}

Finally, let us investigate the phase diagram for an
unrealistically large diquark coupling of $G_D/G_S = 1.5$,
cf.\ Fig.~\ref{rischke_fig:phased3}.
We observe that the BEC region becomes rather wide.
The critical temperatures $T_c$ for the 2SC and CFL phases
and $T_{\rm diss}^c$ also increase substantially.
If the diquark coupling is raised further, even
the vacuum at $T=\mu=0$ becomes a 
Bose-Einstein condensate of diquark molecules.

\section{Conclusions}
\label{rischke_conc}

In this paper, we studied the phase diagram of 
three-flavor quark matter with particular emphasis
on the formation of bound diquark molecules and the possibility
that they undergo Bose-Einstein consensation (BEC).
We found that diquark molecules appear
at low densities and temperatures smaller than
their dissociation temperature $T_{\rm diss}$.
This dissociation temperature may 
exceed the deconfinement temperature and
implies that diquarks can be tightly bound
even in the quark-gluon-plasma phase.
We also found that BEC of diquarks is possible for
large diquark interaction strengths $G_D$. This
finding is in agreement with the results of Ref.\
\cite{Abuki:2006dv}.
However, the required values for $G_D$ are
probably unrealistically large, which makes it rather
unlikely that a BEC of diquarks will be observed
in compact stellar objects or even heavy-ion collisions.

\section*{Acknowledgements}
D.H.R.\ thanks the Yukawa 
Institute for Theoretical Physics at Kyoto University,
and in particular Professor Teiji Kunihiro,
for making his visit a most pleasant one,
and for supporting him through
JSPS fellowship S-06184.


\end{document}